\documentclass[pra,aps,superscriptaddress,showpacs,reprint]{revtex4-1}
\usepackage{amsmath}
\usepackage{amsfonts}
\usepackage{amssymb}
\usepackage{dcolumn}
\usepackage{bm}
\usepackage[dvips]{graphicx}
\usepackage{epsfig}

\begin{document}
\title{Landau-Zener and Rabi oscillations in the spin-dependent conductance}

\author{Lucas J. Fern\'{a}ndez-Alc\'{a}zar}
\affiliation{Instituto de F\'{i}sica Enrique Gaviola (IFEG), CONICET-UNC and Facultad
de Matem\'{a}tica, Astronom\'{i}a y F\'{i}sica, Universidad Nacional
de C\'{o}rdoba, 5000, C\'{o}rdoba, Argentina}

\author{Horacio M. Pastawski}
\email{horacio@famaf.unc.edu.ar}
\affiliation{Instituto de F\'{i}sica Enrique Gaviola (IFEG), CONICET-UNC and Facultad
de Matem\'{a}tica, Astronom\'{i}a y F\'{i}sica, Universidad Nacional
de C\'{o}rdoba, 5000, C\'{o}rdoba, Argentina}

\begin{abstract}
We describe the spin-dependent quantum conductance in a wire where a magnetic
field is spatially modulated. The changes in direction and intensity of the 
magnetic field acts as a perturbation that mixes spin projections.
This is exemplified by a ferromagnetic nanowire. There the local field varies
smoothly its direction generating a domain wall (DW) as described by the well
known Cabrera-Falicov model. Here, we generalize this model to include also a
strength modulation. We identify two striking diabatic regimes that appear
when such magnetic inhogeneity occurs. 1) If the field strength at the DW is
weak enough the local Zeeman energies result in an avoided crossing. Thus, the
spin flip probability follows the Landau-Zener formula. 2) For strong fields,
the spin-dependent conductance shows oscillations as function of the DW
width. We interpret them in terms of Rabi oscillations. Time and length
scales obtained from this simplified view show an excellent agreement with the
exact dynamical solution of the spin-dependent transport. These results remain
valid for other situations involving modulated magnetic structures and thus
they open new prospects for the use of quantum interferences in spin based devices.
\end{abstract}

\maketitle

\section{Introduction} The control and design of spin-dependent electronic transport in
magnetically modulated devices represents a promising technological
challenge.\cite{DaSarmaRMP04} Spintronic devices switch the spin state or
filter electrons by spin. The most direct way to tune the transport, aside
spin-orbit effects,\cite{EguesLossPRL03,UsajEguesPRL08} is to use designed
magnetic inhomogeneities which couple directly to the spin. The most
prominent devices are those based on the giant magnetoresistance.\cite{F08}
Other recent developments involve spin valves based on organic molecules,%
\cite{SteilNature,DedNature} and quasi one-dimensional spin transistors. 
\cite{Betthausen12,Ritch12} Since quantum effects become relevant, transport
is based on the Landauer's motto that \textquotedblleft conductance is
transmittance\textquotedblright .\cite{LImry99}

Previous studies of spin-dependent quantum transport, suggest the presence
of interesting physical phenomena. For example, conductance through
ferromagnetic nanowires with a domain wall (DW) shows
some Fabry-P\`{e}rot like interferences, which were not fully understood.%
\cite{GWJS04,FJWS04} Also, transport on magnetically modulated
semiconducting spin valves\cite{Ritch12} showed magnetic commensurabilities
as well as regimes compatible with a Landau-Zener problem. Thus, quantum
transport through magnetic inhomogeneities becomes a promising tool in
spintronics, where the different characteristic times and lengths
should be identified. These scales should be compared with the electron's
Fermi wavelength. The tunneling adiabaticity is given by the electron's
speed. In this context, a dynamical description of the transport process
would improve the comprehension of these phenomena.

In this letter we consider a variant of the Cabrera and Falicov \cite{CF74}%
model for spin dependent electronic transport through a soft magnetic DW. It
is representative of a wide class of magnetic inhomogeneities. In the
original model the field just rotates along the DW. We extend it allowing a
modulation in the field strength. This simple variation will have non
trivial consequences on transport. We show that in a weak perturbation
regime, the spin-dependent conductance through the DW can be described by
the Landau-Zener (LZ) formula. In contrast, for a strong perturbation
regime, we find well defined interferences as a function of the DW width,
which are interpreted as Rabi oscillations. This interpretation is
confirmed by an analysis of the wave packet dynamics. The physics and
computational strategies described here could help in the design of better
spintronic devices.

\section{Hamiltonian of the conduction electrons}

We consider a single spin channel:\cite{GWJS04} 
\begin{equation}
\hat{H}=-\frac{\hbar ^{2}}{2m}\frac{\mathrm{d}^{2}}{\mathrm{d}x^{2}}-\vec{\mu%
}\cdot \vec{B}(x).
\end{equation}%
The first term is the\ kinetic energy along $x$ of electrons with effective
mass $m$. The second term is the Zeeman interaction between the spin
magnetic moment $\vec{\mu}$ and $\vec{B}(x),$ the effective magnetic field
at $x$. Here, $\vec{\mu}=-\mu _{B}\vec{\sigma}$ , where $\mu _{B}$ is the
Bohr magneton and $\vec{\sigma}=(\sigma _{x},\sigma _{y},\sigma _{z})$ is
the vector of the Pauli matrices. In particular, the dependence on $x$ of
the magnitude and direction of $\vec{B}(x)$ may cause the spin-dependent
scattering. These inhomogeneities may be natural, as in ferromagnetic DWs,%
\cite{Czer08} or artificially generated, as in magnetic semiconducting
waveguides. \cite{Betthausen12} We will express our results in the concrete
language of ferromagnetic nanowires.%

\begin{figure*}[ptb]
\begin{center}
\includegraphics[width=0.8\textwidth]%
{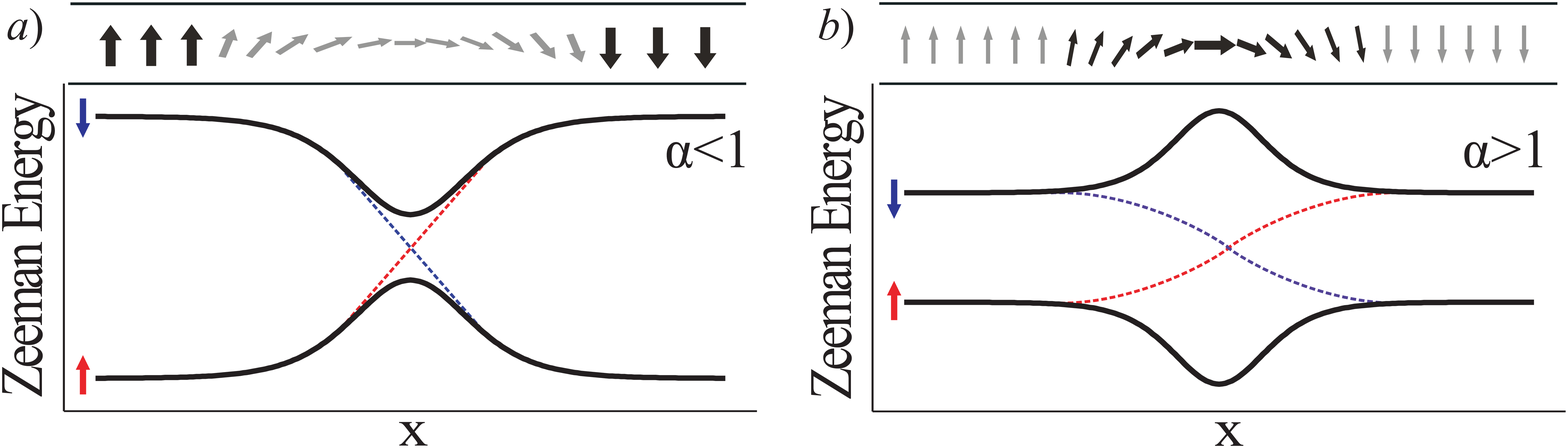}%
\caption{(Color online) Solid black lines are the local eigenenergies for
electrons with $\uparrow$ and $\downarrow$ spin in presence of a locally
rotated field which is schematized on top. $a)$ weak ($\alpha<1$) or $b)$
strong ($\alpha>1$) field strength at the DW center. Dashed lines are the
unperturbed Zeeman energies.}%
\label{fig_Zeemanlocal}%
\end{center}
\end{figure*}

\section{Electronic structure in modulated magnetic fields}

The Cabrera and Falicov soft-DW model\cite{CF74} considers a quantum spin
channel laid along $x$, and a magnetic field whose orientation rotates as it
progress along the DW. We generalize this description by including a
modulation in the field intensity. The vector $\vec{f}(x)=\left(
f_{x}(x),f_{y}(x),f_{z}(x)\right) =-\vec{B}(x)/B_{\infty }$, %
describes the DW shape, and satisfies $\left\vert \vec{f}(x)\right\vert
\rightarrow 1$ as $x\longrightarrow \pm \infty $. The asymmetry in the
modulation strength is described by the parameter $\alpha =B(0)/B_{\infty }$%
. Using the axis $z$ as quantization direction for the spin, 
\begin{equation}
\vec{f}(x)=\left( \alpha /\cosh (\frac{x}{W})~,~0~,~\tanh (\frac{x}{W}%
)\right) .
\end{equation}%
Here, $W$ is the half-width of the DW. Obviously, $\alpha =1$ corresponds
to a field of constant strength that rotates across the DW. The
Zeeman term is $\vec{\mu}\cdot \vec{B}(x)=\Delta _{0}\vec{\sigma}\cdot \vec{f%
}(x)$. Here, $\Delta _{0}=\mu _{B}B_{\infty }$. The wave function for a
conduction electron has components along both spin directions, $\left\vert
\uparrow \right\rangle $ and $\left\vert \downarrow \right\rangle $,
referred to the quantization axis parallel to the field at the left domain
(laboratory frame), as 
\begin{equation}
\left\vert \psi (x)\right\rangle =\varphi _{\uparrow }(x)\left\vert
x,\uparrow \right\rangle +\varphi _{\downarrow }(x)\left\vert x,\downarrow
\right\rangle .
\end{equation}%
Therefore, the equations governing the electron tunneling and the spin flip
are:%
\begin{equation}
\left\{ 
\begin{array}{c}
-\dfrac{\hbar ^{2}}{2m}\dfrac{\mathrm{d}^{2}}{\mathrm{d}x^{2}}\varphi
_{\uparrow }(x)+E_{\uparrow }(x)\varphi _{\uparrow }(x)+V_{\uparrow
\downarrow }(x)\varphi _{\downarrow }(x)=\varepsilon \varphi _{\uparrow }(x)
\\ 
\\ 
-\dfrac{\hbar ^{2}}{2m}\dfrac{\mathrm{d}^{2}}{\mathrm{d}x^{2}}\varphi
_{\downarrow }(x)-E_{\downarrow }(x)\varphi _{\downarrow }(x)+V_{\uparrow
\downarrow }(x)\varphi _{\uparrow }(x)=\varepsilon \varphi _{\downarrow }(x),%
\end{array}%
\right.   \label{eq_principal}
\end{equation}%
where $\varepsilon $\textbf{\ }is the energy associated with the dynamics
along $x$, identifying,%
\begin{align}
\mu _{B}B_{z}(x)& =E_{\downarrow }(x)=\Delta _{0}\tanh (\frac{x}{W}),~\text{%
and}  \label{eq_energy} \\
\mu _{B}B_{x}(x)& =V_{\uparrow \downarrow }(x)=\alpha \Delta _{0}/\cosh (%
\frac{x}{W}),
\end{align}%
where $E_{\downarrow }(x)=-E_{\uparrow }(x)$. The local states become mixed
by $V_{\uparrow \downarrow }(x)$ while the electron moves through the DW.

\section{Evaluation of the conductance}

To evaluate the quantum conductance we will use the Landauer-B\"{u}ttiker
equation. \cite{Buttiker-multichannel} There, different conductances are
given by the transmittances between states of definite momentum and spin
projection at the contacts .\cite{LImry99}

In a tight-binding representation,\cite{PM01} the spatial coordinate takes
discrete values $x_{n}$ in a grid of unit $a$, $x_{n}\rightarrow na$. Every
site in the grid has an associated normalized local wave function $%
\left\vert n\right\rangle ,$ which will be called $n^{\mathrm{th}}$ \textit{%
orbital} as in a LCAO scheme. Each orbital has an energy given by the local
potential $E_{s}(x_{n})=E_{n,s}$, where $s$ is either $\uparrow $ or $%
\downarrow $, and the transverse field yields $V_{\uparrow \downarrow
}(x_{n})=V_{n,\uparrow \downarrow }$. Any electronic wave function with well
defined spin $s$ is now written in terms of a discrete sum: 
\begin{equation}
\left\vert \varphi _{s}\right\rangle \rightarrow \sum_{n}u_{n,s}\left\vert
n,s\right\rangle ,
\end{equation}%
where, according to eq. \ref{eq_principal}, the spin-orbital amplitudes $%
u_{n,s}$ must satisfy:%
\begin{equation}
\left\{ 
\begin{array}{c}
-V\left[ u_{n+1,\uparrow }-2u_{n,\uparrow }+u_{n-1,\uparrow }\right]
+E_{n,\uparrow }u_{n,\uparrow } \\ 
\phantom{~~\varepsilon u_{n,\uparrow}+E_{n,\uparrow}u_{n,\uparrow} }%
+V_{n,\uparrow \downarrow }u_{n+1,\downarrow }=\varepsilon u_{n,\uparrow }
\\ 
\\ 
-V\left[ u_{n+1,\downarrow }-2u_{n,\downarrow }+u_{n-1,\downarrow }\right]
+E_{n,\downarrow }u_{n,\downarrow } \\ 
\phantom{~~~\varepsilon u_{n,\uparrow}+E_{n,\uparrow}u_{n,\uparrow} }%
+V_{n,\uparrow \downarrow }u_{n+1,\uparrow }=\varepsilon u_{n,\downarrow }.%
\end{array}%
\right.
\end{equation}

The unit of energy is given by the hopping strength $V=\hbar ^{2}/(2ma^{2})$%
. We consider energies at the band center ($\varepsilon \simeq 2V$). Each
spin orientation is represented by a chain with $N\gg 1$ orbitals that
comprises the whole DW. $L$ and $R$ are sites indices symmetrically arranged
at the left and right sides of the DW and satisfying $(R-L)a=Na\gg W$. Then $%
E_{n,s}\equiv E_{L,s}=\pm \Delta _{0}$ for $n\leq L$ and $E_{n,s}\equiv
E_{R,s}=\mp \Delta _{0}$ for $n\geq R$. Since far away of the DW, $%
V_{\uparrow \downarrow }\equiv 0$, the asymptotic eigenvalues, with wave
vector $k,$ are $\varepsilon _{k,s}=\pm \Delta _{0}+2V-2V\cos \left(
ka\right) $. In the region of the DW both spin orientations become coupled
by the perpendicular component of the magnetic field, represented by the\
hopping element $V_{n,\uparrow \downarrow }$.

The magnetic domains, which play the role of contacts, are described through
a renormalization procedure. \cite{LevsteinPastawskiDAmato} In an open
system it leads to a non-Hermitian effective Hamiltonian:\cite{P07}%
\begin{equation}
\hat{H}_{T}=\hat{H}+\hat{\Sigma},
\end{equation}%
where

\begin{align}
\hat{\Sigma}(\varepsilon )& =\Sigma _{L\uparrow }(\varepsilon )\left\vert
L\uparrow \right\rangle \left\langle L\uparrow \right\vert +\Sigma
_{L\downarrow }(\varepsilon )\left\vert L\downarrow \right\rangle
\left\langle L\downarrow \right\vert   \notag \\
& +\Sigma _{R\uparrow }(\varepsilon )\left\vert R\uparrow \right\rangle
\left\langle R\uparrow \right\vert +\Sigma _{R\downarrow }(\varepsilon
)\left\vert R\downarrow \right\rangle \left\langle R\downarrow \right\vert .
\end{align}%
$\Sigma _{j}$ are the self-energies that satisfy the Dyson equation in the
magnetic domains:%
\begin{align}
\Sigma_{j}(\varepsilon)  &  =\frac{V^{2}}{\varepsilon-E_{j}-\Sigma
_{j}(\varepsilon)}=\operatorname{Re}\Sigma_{j}(\varepsilon)-\mathrm{i}%
\Gamma_{j~}(\varepsilon)\\
&  \simeq-\mathrm{i}\Gamma_{j},\text{ at the band center.}%
\end{align}
The double spin-orbital subscript $j=L\uparrow ,~L\downarrow ,~R\uparrow $
or $~R\downarrow $ indicates the left ($L$) or right ($R$) channels inside
the magnetic domains with the corresponding spin orientation.  $2a\Gamma
_{j}/\hbar $ is the group velocity at the spin channels\textbf{\ }$j$\textbf{%
\ } connected to each spin-orbitals at the sides, \textit{i.e.} $L\uparrow
,~L\downarrow ,~R\uparrow $ and $~R\downarrow $ in a four-terminal circuit.

We obtain the retarded and advanced Green functions from $\hat{H}_{T}$ as $%
\hat{G}^{R}(\varepsilon )=\lim_{\eta \rightarrow 0^{+}}\left( \varepsilon +%
\mathrm{i}\eta -\hat{H}_{T}\right) ^{-1}$  and $\hat{G}^{A}=\hat{G}%
^{R\dagger }$. The transmittance is: \cite{PM01}%
\begin{equation}
T_{ij}(\varepsilon )=2\Gamma _{i}(\varepsilon )\left\vert
G_{ij}^{R}(\varepsilon )\right\vert ^{2}2\Gamma _{j}(\varepsilon ).
\label{eq_transmitancia_FisherLee}
\end{equation}%
Here, $\Gamma _{j}=\mathrm{Im}(\Sigma _{j})$ and $i,j=L\uparrow
,~L\downarrow ,~R\uparrow ,~R\downarrow $, being $j$ and $i$ the electronic
input and output spin-orbital channels respectively. When the evaluated
channels correspond to opposite spin projections in opposite sides of the DW
we call them spin-flip transmittances, e.g. $T_{\downarrow \uparrow }$. 

\section{Transport in the regime of $\protect\alpha<1$ as a Landau-Zener
problem}

In eq. \ref{eq_principal} $V_{\uparrow \downarrow }$ is responsible for the
mixture of the spin orientations. While $V_{\uparrow \downarrow }(x)$
vanishes within the domains, it is roughly constant at the DW center.
Besides, $E_{\uparrow }$ and $E_{\downarrow }$ account for the Zeeman energy
in the laboratory frame and they have the meaning of effective potentials
for those electrons oriented along the field. While, inside the domains, $%
E_{\uparrow }-E_{\downarrow }=$ $2\Delta _{0}$ quantifies the Zeeman
splitting, at the DW center both energies intersect $E_{\uparrow
}=E_{\downarrow }=0$.
\begin{figure}
[ptb]
\begin{center}
\includegraphics[width=0.4\textwidth]%
{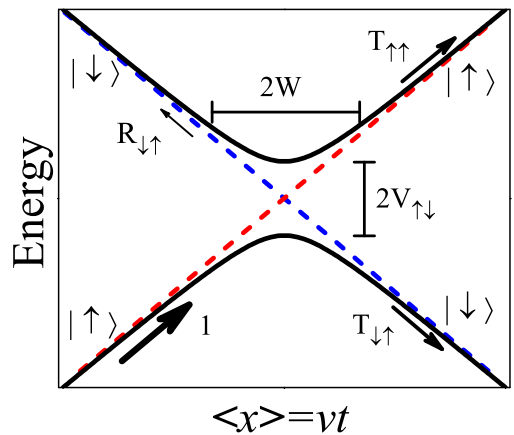}%
\caption{(Color online) Scheme of the eigenenergies of a two level system,
$\left\vert \uparrow\right\rangle $ and $\left\vert \downarrow\right\rangle $,
as functions of the dynamical parameter, $x=vt,$ that controls the
Landau-Zener transition. The energy levels show an avoided crossing due to the
presence of a perturbation that couples the states. The unperturbed energies
are shown with dashed lines. Energy and coordinate have arbitrary dimensions.}%
\label{fig_landau_zener}%
\end{center}
\end{figure}

Consider an electron wave packet with a given spin projection that moves
with definite momentum towards the DW. Its\ mean position results
proportional to the elapsed time $\left\langle x(t)\right\rangle \equiv
x\simeq v_{F}t$. The spin dependent mean potential energy will change as the
wave packet moves through the DW and starts to be mixed with that of the
opposite spin by the $V_{\uparrow \downarrow }$ term. These potential
energies are assimilable to the time dependent local energies of a two level
system in a LZ problem.\cite{Zener32} This last describes the transition
between two states when their unperturbed energies are swept across a
degeneracy point while a constant perturbation produces an avoided crossing.
In fig. \ref{fig_landau_zener} we show the energy diagram for the DW region.
The comparison between this diagram and the local Zeeman levels, in fig. \ref%
{fig_Zeemanlocal}$a)$,\ evidences the analogy between the LZ problem and the 
$\alpha <1$ regime. Thus, the LZ equation predicts that the probability to
exit in the state $\left\vert \downarrow \right\rangle $ to the right,
provided that it entered from the left in the state\ $\left\vert \uparrow
\right\rangle ,$ is:%
\begin{equation}
P_{\downarrow \uparrow }=1-\exp \left( -\frac{2\pi }{\hbar }\frac{\left\vert
V_{\uparrow \downarrow }\right\vert ^{2}}{(\mathrm{d}E(0)/\mathrm{d}t)}%
\right) .
\end{equation}%
Here, we can use $E(x)=E_{\uparrow }(x)-E_{\downarrow }(x)$ and $x\simeq
v_{F}t$ to evaluate the derivative. The adiabaticity parameter $2\pi
\left\vert V_{\uparrow \downarrow }\right\vert ^{2}/(\hbar \mathrm{d}E/%
\mathrm{d}t)$ describes a fully adiabatic transition if it is much greater
than $1$ resulting in $P_{\downarrow \uparrow }\lesssim 1$, while the
opposite limit is a diabatic process where $P_{\downarrow \uparrow }\gtrsim
0 $.

We will consider a wave packet with  $\varepsilon _{F}\gg \Delta _{0}$ and $%
v_{F}(x)\simeq 2aV/\hbar $. Hence, the time of transit through the DW is $%
\tau _{W}=2W/v_{F}$. We can relate the DW crossing\textbf{\ }with the LZ
problem identifying: 
\begin{align}
\frac{\mathrm{d}E}{\mathrm{d}t}& =\frac{\mathrm{d}E}{\mathrm{d}x}\frac{%
\mathrm{d}x}{\mathrm{d}t}  \notag \\
& \simeq \frac{2\Delta _{0}}{W}v_{F},  \label{eq_dE_sobre_dt}
\end{align}%
where $E(x)$ is calculated as in the LZ formula, being $E_{\uparrow }$ and $%
E_{\downarrow }$\ obtained from eq. \ref{eq_energy}. Thus, the adiabaticity
parameter results $\pi \left( a/W\right) \left\vert V_{\uparrow \downarrow
}\right\vert ^{2}/\left( \Delta _{0}V\right) ,$ where $V_{\uparrow
\downarrow }=\alpha \Delta _{0}$. We will choose to control the adiabaticity
of the crossing by changing $W$.

\section{Numerical Results}

\begin{figure}
[ptb]
\begin{center}
\includegraphics[
height=3.5094in,
width=3.0424in
]%
{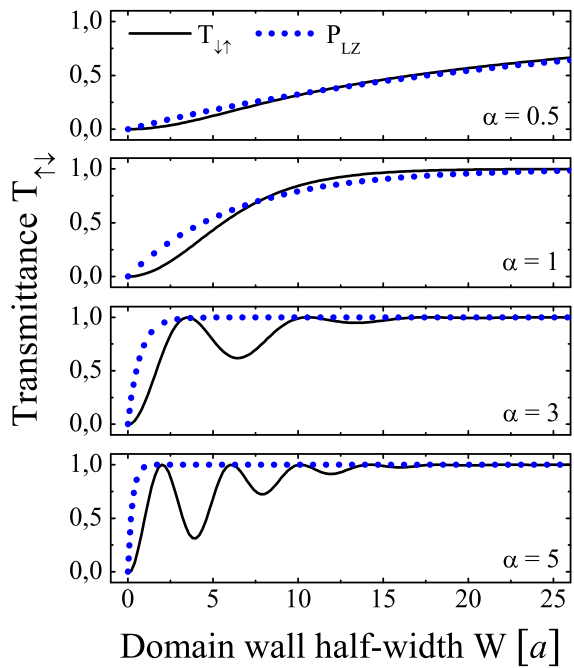}%
\caption{(Color online) The transmittance $T_{\downarrow\uparrow}$ versus the
DW half-width $W$ for different $\alpha.$ The energy is measured in units of
$V$. The Zeeman splitting is $2\Delta_{0}=0.2V$ and $V_{\uparrow\downarrow
}=\alpha\Delta_{0}$. The transition probabilities predicted by the
Landau-Zener formula $P_{LZ}$\ are shown with dashed lines.}%
\label{fig_trans}%
\end{center}
\end{figure}

We will use eq. \ref{eq_transmitancia_FisherLee} to evaluate the
spin-dependent transmittance, $T_{\downarrow\uparrow}$, that describes the
spin-flip process. In fig. \ref{fig_trans} we compare $T_{\downarrow\uparrow}
$ with the spin-flip probability of the LZ problem. Both are shown as
function of the parameter $W$. We show four different DW characterized by $%
\alpha=0.5,~1,~3~$and$~\ 5$. In all cases we consider $\Delta_{0}=0.1$ and $%
V=1$.

The two upper panels of fig. \ref{fig_trans} may be associated to $\alpha
\leq 1$ regime. At the center of the DW, $\left. V_{\uparrow \downarrow
}(x)\right\vert _{x=0}=V_{\uparrow \downarrow }^{\max }=0.5\Delta _{0}$ and $%
V_{\uparrow \downarrow }^{\max }=1\Delta _{0}$. If the DW is abrupt, \textit{%
i.e.} $W\simeq 0$, the electrons keep their initial spin orientation, thus $%
T_{\downarrow \uparrow }\simeq 0$. This coincides with the regime where the
giant magnetoresistance arises from the scattering at the DW.\cite{F08} On
the other hand, if the DW width is broad enough, the electrons tend to
change their spin orientation $T_{\downarrow \uparrow }\simeq 1$, preventing
the magnetoresistance. Considering the overall dependence on $W$ for $\alpha
\leq 1$, we appreciate that there is a fair correspondence between the
transmittance and the LZ probability that improves as $\alpha $ becomes
smaller. Even when there are small discrepancies, these can be attributed to
the complexity inherent to our problem, where the \textquotedblleft
perturbation\textquotedblright\ $V_{\uparrow \downarrow }$ gradually turns
on while the levels become degenerate at the DW center. This exceeds the
simplicity of LZ model.

The two lower panels\ of fig. \ref{fig_trans} correspond to a $\alpha>1$
regime, with $V_{\uparrow\downarrow}^{\max}=3\Delta_{0}$ and $V_{\uparrow
\downarrow}^{\max}=5\Delta_{0}$. In both cases the transmittances oscillate
as function of $W$. This might suggest an analogy to the Fabry-P\'{e}rot
interferences in tunneling problems. In the case of ref. \cite{GWJS04} this
phenomenon is justified because their potentials have slopes with
discontinuities at the DW. However, here we consider a high-energy problem
with smooth potential barriers. Therefore, it always results $T_{\downarrow
\uparrow}+T_{\uparrow\uparrow}\lesssim1$ and the reflectances are nearly
zero. Hence, the Fabry-P\'{e}rot interferences are discarded as possible
origin of the observed oscillations. Instead, while the electron moves
across the DW its spin oscillates between the states $\left\vert
\uparrow\right\rangle $ and $\left\vert \downarrow\right\rangle $ driven by
the \textquotedblleft perturbation\textquotedblright\ $V_{\uparrow%
\downarrow} $. This is consistent with the fact the spin precesses around
the local field. This is called Larmor precession for a semiclassical spin
and Rabi oscillation for a spin-1/2.\cite{Rabi_Oscillation} This $\pi\left(
a/W\right) \left\vert V_{\uparrow\downarrow}\right\vert ^{2}/\left(
\Delta_{0}V\right) \lesssim1$ regime, contrasts with the adiabatic
transition where the electron's spin simply remain aligned with the local
magnetic field while it crosses the DW. However, since both DWs are smooth,
the oscillation frequency varies continuously and thus it is not obvious
that well defined Rabi oscillations would show up.

\section{Transport in the regime of $\protect\alpha>1$: interferences as
Rabi oscillations}

In the DW, the Zeeman energies in the laboratory frame are degenerate while
the coupling $V_{\uparrow \downarrow }(x)$  is maximum $V_{\uparrow
\downarrow }=\left. V_{\uparrow \downarrow }(x)\right\vert _{x=0}=\alpha
\Delta _{0}$. Locally, this can be seen as a two level system undergoing a
Rabi oscillation with period 
\begin{equation}
\tau _{R}=\dfrac{\pi \hbar }{V_{\uparrow \downarrow }}.  \label{ec_T_rabi}
\end{equation}%
Therefore, the length traveled by the electron during that Rabi cycle is 
\begin{equation}
L_{R}=\tau _{R}v_{F}=2\dfrac{\pi aV}{\alpha \Delta _{0}},  \label{ec_L_rabi}
\end{equation}%
where $v_{F}\simeq 2aV/\hbar $. We adopt the term \textquotedblleft Rabi
oscillation\textquotedblright\ to emphasize that the spin-1/2 is in an
oscillating superposition of its two possible projections.

We analyze the spin-flip transmittances in terms of the length scales
estimated above. In the case of the two upper panels of fig. \ref{fig_trans}
, $\alpha=0.5$ and $\alpha=1,$ the Rabi oscillation might have, according to
eq. \ref{ec_L_rabi}, characteristic lengths of $L_{R}\simeq125.6a$ and $%
L_{R}\simeq62.8a$ respectively. These are much longer than the DW width
needed for an adiabatic spin-flip. This explains the absence of oscillations
and the applicability of the LZ formula. In contrast, the $\alpha=3$ and $%
\alpha=5$ cases, shown in the two lower panels, the spin-flip transmittances
present oscillations with characteristic lengths of $L_{N}\simeq8a$ and $%
L_{N}\simeq4a,$ respectively. According to our hypothesis of transmittances
modulated by Rabi oscillations, the spacing between two consecutive local
minima must be $L\simeq10.5a$ and $L\simeq6.3a$, respectively. The
discrepancy between our na\"{\i}ve prediction and the numerical results are
justified by the fact that the Rabi length is not a perfectly defined
magnitude in our\ smooth DW model. This is because $V_{\uparrow\downarrow},$
and hence the involved periods, change as the electron moves through the DW. 
\begin{figure}
[h]
\begin{center}
\includegraphics[width=0.48\textwidth
]%
{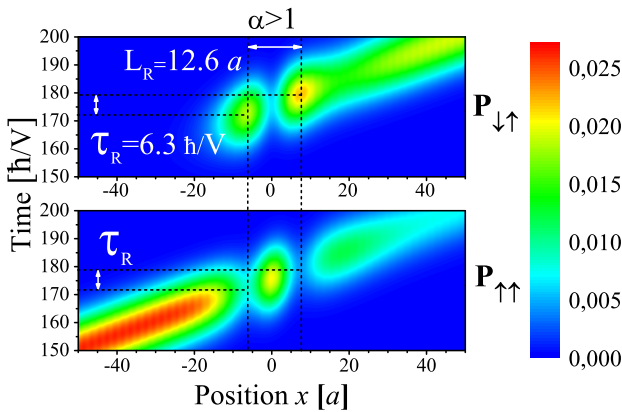}%
\caption{(Color online) Probability (color scale) for a $\downarrow$ spin
orientation (upper panel) or $\uparrow$ spin orientation (lower panel) as
function of time and position, given that the initial electron state at the
left has $\uparrow$ spin and moves with the Fermi velocity $v_{F}=2aV/\hbar.$
Here, $\Delta_{0}=0.1V$, $W=8a$ and $\alpha=5$. The midpoint of the DW is
located at $x=0a$. The probability oscillations confirm that Rabi
oscillations are present. The vertical and horizontal short-dashed lines
come into contact at the maxima and minima of the oscillations. From these,
the magnitudes of the period and oscillation characteristic length can be
inferred. These coincide with those predicted by the Eqs. \ref{ec_T_rabi} and
\ref{ec_L_rabi}, $\tau_{R}=6.3\hbar/V$ and $L_{R}=12.6a$ as shown by the white
arrows.}%
\label{fig_dinamica_Rabi1}%
\end{center}
\end{figure}
\begin{figure}
[h]
\begin{center}
\includegraphics[width=0.48\textwidth
]%
{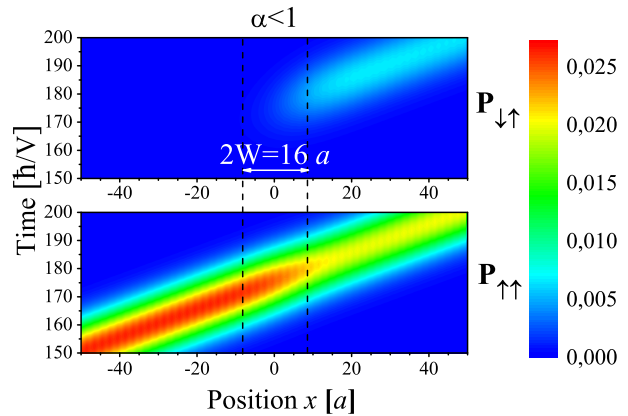}%
\caption{(Color online) Probability (color scale) for a $\downarrow$ spin
orientation (upper panel) and $\uparrow$ spin orientation (lower panel) as
function of time and position, for $\Delta_{0}=0.1V$, $W=8a$ and $\alpha=0.5$.
\ The dashed lines indicate the effective width of the DW, $2W$. The Rabi
oscillations are not developed. }%
\label{fig_dinamica_Rabi2}%
\end{center}
\end{figure}

In order to confirm the Rabi oscillation hypothesis, we analyze the
dynamical evolution of the electron's spin implementing an unitary algorithm
based on the Trotter approximation.\cite{trotter} We consider an initial
Gaussian wave packet with spin $\left\vert \uparrow \right\rangle $ and a
well defined momentum at the band center. This last condition avoids
undesired effects of dispersion. During the electron's transit through the
DW, the spin projection tries to follow the magnetic field and thus the
final spin projection depends on $W$.

In fig. \ref{fig_dinamica_Rabi1} we show the probability densities $%
P_{\downarrow\uparrow}(x,t)$ and $P_{\uparrow\uparrow}(x,t)$ for $\alpha=5$.
Here, the second subscript is the initial spin, while the first one
indicates the corresponding\ spin projection at time $t$. The upper panel
shows the probability for the $\left\vert \uparrow\right\rangle $ spin
projection, while the lower panels show the $\left\vert
\downarrow\right\rangle $ spin projection. The Zeeman splitting inside the
domains is $2\Delta_{0}$, with $\Delta_{0}=0.1V$. The DW center is placed at 
$x=0a$ and its width is $2W=16a,$ which implies an adiabaticity parameter
of about $1$. According to fig. \ref{fig_trans}, this $W$ ensures some
oscillations in the transmittance. The intensity plot is consistent with the
fact that as the electron moves through the DW, the probability of finding
the electron with the spin up projection decreases while the complementary
spin down density increases. Subsequently, an increase in the spin up
probability is produced while a decrease occurs for the opposite spin
projection. This cycle is repeated until the electron reaches the end of the
DW. This oscillation between the two spin projections is identified with a
Rabi oscillation. The observed period and characteristic length are in full
agreement with those given by Eqs. \ref{ec_T_rabi} and \ref{ec_L_rabi}, $%
\tau_{R}=6.3\hbar/V$ and $L_{R}=12.6a$ respectively. These magnitudes are
drawn in fig. \ref{fig_dinamica_Rabi1}.

The same analysis is performed for a $\alpha=0.5$ DW and shown in fig. \ref%
{fig_dinamica_Rabi2}. This $\alpha,$ together with $2W=16a,$\ implies a low
adiabaticity parameter of about $0.01$. Again, the final electronic state $%
\left\vert \varphi_{R}\right\rangle $ at the right domain, after traversing
the DW, is a superposition of the two spin projections. The probability of
finding the state $\left\vert \downarrow\right\rangle $, is consistent with
the transmittance shown in fig. \ref{fig_trans} and the LZ prediction: $%
T_{\downarrow\uparrow}\equiv~\left\vert \left\langle \downarrow\right.
\left\vert \varphi_{R}\right\rangle \right\vert ^{2}~=P_{\downarrow\uparrow}$
and $T_{\uparrow\uparrow}=~\left\vert \left\langle \uparrow\right.
\left\vert \phi_{R}\right\rangle \right\vert
^{2}=P_{\uparrow\uparrow}~\simeq 1-T_{\downarrow\uparrow}$. As consequence
of the DW smoothness there are no significative reflections. In contrast to
the previous case, we see that the time oscillations are not developed.

While in the present work we just analyzed a single incoming wave vector,
considering a metallic wire would involve integrating, up to the Fermi
energy, over transversal channels equivalent to those as described here.
This could smear out the Rabi oscillations reported here and a one would
need a proper design to overcome this difficulty. However, in magnetic
semiconducting waveguides, the relevant role of lateral quantization leaves
the considered model as a realistic description.\cite{Ritch12}\textbf{\ }

\section{Conclusion}

In this letter we explored the quantum phenomena associated to
spin-dependent transport in presence of a smooth magnetic inhomogeneity,
much as a DW in a magnetic nanowire. For this purpose, we extended the
Cabrera-Falicov model to account for modulations on the magnetic field
intensity at the DW. The physics we described is not restricted to this
case.\ Indeed, our results and strategies remain valid for other situations,
such as magnetically modulated semiconducting structures. \cite{Betthausen12}
There, spectral modulations are described by variants of eq. \ref%
{eq_principal}. In ref. \cite{Ritch12} a situation assimilable to our $%
\alpha <1$ is presented for a spin transistor based on helical magnetic
fields.

We showed that, for $\alpha \leq 1,$ the spin dependent transport across the
magnetically modulated region are fairly described by the LZ formula. LZ
applies to the whole dynamical range, from diabatic to fully adiabatic
crossing. We showed that, by performing the appropriate mapping of the
relevant variables, LZ yields a quite fair description under well defined
conditions for the perturbation. For $\alpha >1$, we found that conductance
has quantum interferences which manifest as oscillations as function of the
DW width. These can not be assigned to Fabry-P\'{e}rot interferences. By
performing a dynamical study of the tunneling process, we showed that a spin
polarized wave packet propagating across the DW, can be seen as a two level
system undergoing Rabi oscillations.

A possible experimental set up to test these effect in all regimes could be
a linear semiconducting waveguide in presence of a locally modulated field.
In such a case, few conducting channels are enabled by a gate voltage that
also controls the carrier's wavelength. Finally, the dynamical description
of the transport problem as presented here, may prove useful for the
converse problem: \textit{i.e.} evaluating the dynamics of a DW under pulsed
electrical currents. This may extend the interest of our strategy to study a
problem of growing interest: electrically driven domain-wall-based memories
in quasi one-dimensional (1D) magnetic wires.\cite{Parkin07}

In summary, for the perturbative regime ($\alpha <1$), we probed a definite
connection between steady-state spin-dependent transport across a
magnetically modulated region and the time dependent Landau-Zener problem.
In the strong perturbation regime ($\alpha >1$), we showed that the
steady-state conductance presents interferences. We probed that they arise
from Rabi oscillations, by performing a time dependent calculation.

\acknowledgments
HMP wants to dedicate this work to the memory of
his life long collaborator Patricia Rebeca Levstein. This work was performed
with the financial support from CONICET, ANPCyT, SeCyT-UNC and MinCyT-Cor.

\end{document}